\documentstyle[fleqn]{tp}
\input psfig

\def\etal{{\it et\thinspace al.\/~}}

\def\beq#1{\begin{equation}\label{#1}}
\def\eeq{\end{equation}}
\def\beqa#1{\begin{eqnarray}\label{#1}}
\def\eeqa{\end{eqnarray}}

\def\tento#1{\times 10^{#1}}

\def\Ms{\ M_{\odot}}     

\def\K{{\rm \ K}}

\def\cm{{\rm \ cm}}
\def\eV{{\rm \ eV}}

\def\kpc{{\rm \ kpc}}
\def\pc{{\rm \ pc}}

\def\yrs{{\rm \ years}}

\def\HH{H$_2$ }
\def\H2p{H$_2^+$ }
\def\Hp{H$^+$ }
\def\Hm{H$^-$ }
\def\Hep{He$^+$ }
\def\Hepp{He$^{++}$ }

\def\HH{H$_2$ }
\def\H2p{H$_2^+$ }
\def\HHp{H$_2^+$ }
\def\Hp{H$^+$ }
\def\Hm{H$^-$ }
\def\Hep{He$^+$ }
\def\Hepp{He$^{++}$ }

\def\mH2p{H_2^+}

\def\gtsima{$\; \buildrel > \over \sim \;$}
\def\ltsima{$\; \buildrel < \over \sim \;$}
\def\prosima{$\; \buildrel \propto \over \sim \;$}
\def\gsim{\lower.7ex\hbox{\gtsima}}
\def\lsim{\lower.7ex\hbox{\ltsima}}
\def\simgt{\lower.7ex\hbox{\gtsima}}
\def\simlt{\lower.7ex\hbox{\ltsima}}
\def\simpr{\lower.7ex\hbox{\prosima}}

\begin{document}

\twocolumn[
\title{The Formation and Fragmentation of\\ Primordial Molecular Clouds}
\author{Tom Abel$^{1,2}$, Greg Bryan$^{2,3}$ and Michael L. Norman$^{2,4}$\\
{\it $^1$Max-Planck-Institut f\"ur Astrophysik, D--85740 Garching}\\
{\it $^2$LCA, NCSA, University of Illinois, US--61801 Urbana/Champaign}\\
{\it $^3$Massachusetts Institute of Technology,  MA, US--02139 Cambridge}\\
{\it $^4$Astronomy Department, University of Illinois, Urbana/Champaign} }
\vspace*{16pt}   

ABSTRACT.\ Many questions in physical cosmology regarding the thermal
history of the intergalactic medium, chemical enrichment,
reionization, etc.  are thought to be intimately related to the nature
and evolution of pregalactic structure.  In particular the efficiency
of primordial star formation and the primordial IMF are of special
interest.  We present results from high resolution three--dimensional
adaptive mesh refinement simulations that follow the collapse of
primordial molecular clouds and their subsequent fragmentation within
a cosmologically representative volume. Comoving scales from 128 kpc
down to 0.5 pc are followed accurately.  Dark matter dynamics,
hydrodynamics and all relevant chemical and radiative processes
(cooling) are followed self-consistently for a cluster normalized CDM
structure formation model.  Primordial molecular clouds with $\sim
10^5$ solar masses are assembled by mergers of multiple objects that
have formed hydrogen molecules in the gas phase with a fractional
abundance of $\lsim 10^{-4}$.  As the subclumps merge cooling
decreases the temperature to $\sim200$ Kelvin in multiple ``cold
pockets'' of the merger product. Within these cold pockets,
quasi--hydrostatically contracting cores with masses $\sim 100\Ms$ and
number densities $\gsim 10^5 \cm^{-3}$ are found.  We find that
less than 1\% of the primordial gas in such small scale structures
cools and collapses to sufficiently high densities to be available for
primordial star formation.  Furthermore, our results indicate that the
formation of very massive objects, massive black holes, fragmentation
of a large fraction of baryons into brown dwars or Jupiter size
fragments seems, in contrast to various claims in the literature, very
unlikely.  The expected escape fraction of UV photons with
$h\nu<11\eV$ is very small.  \endabstract]

\markboth{Tom Abel et al.}{Primordial Molecular Clouds}

\small

\section{Introduction}

Saslaw and Zipoy (1967) realized the importance of gas phase \HH
molecule formation in primordial gas for the formation of
proto--galactic objects. Employing this mechanism in Jeans unstable
clouds, Peebles and Dicke (1968) formulated their model for the
formation of primordial globular clusters. Further pioneering studies
in this subject were carried out by Takeda \etal (1969), Matsuda \etal
(1969), and Hirasawa \etal (1969) who followed in detail the gas
kinetics in collapsing objects and studied the possible formation of
very massive objects (VMO's).  Hutchins (1976) then looked in greater
detail at the effects of rotation and asked what minimum Jeans mass
can be reached in a collapsing primordial gas cloud. In the 1980's
major contributions to this field were made by Rees and Kashlinsky
(1983), Carr \etal (1984), and Couchman and Rees (1986), who all
studied the possible cosmological consequences of population III star
formation. Their main conclusion was that for hierarchical structure
formation scenarios the first objects might reheat and reionize the
universe and thus raise the Jeans mass, influencing subsequent
structure formation quite dramatically. Massive Pop III stars would
also pre--enrich the intergalactic medium with metals.

It is clear that the uncertainties mentioned above arise due to the
inherently multidimensional, nonlinear, nonequilibrium physics which
determine the collapse and possible fragmentation of gravitationally
and thermally unstable primordial gas clouds. The computational
expense of solving the network of chemical rate equations forced early
studies to focus on single cell calculations (cf.  Hirasawa 1969;
Hutchins 1976; Palla \etal 1983; MacLow and Shull 1986; Puy \etal
1996; Tegmark \etal 1997) adopting simple collapse models.
Bodenheimer (1986) was the first to address the hydrodynamic aspects
of the problem, in spherical symmetry.  Similarly, using a spherical
Lagrangian hydrodynamics code and solving the kinetic rate equations
simultaneously, Haiman, Thoul and Loeb (1996) studied the important
question of which mass scales are able to cool efficiently enough to
collapse. Unfortunately, the issue of fragmentation cannot reliably be
addressed in such spherically symmetric models.

Multi--dimesional studies of first structure formation have only
recently become computationally feasible (Abel 1995, Anninos \& Norman
1996, Zhang \etal1997, Gnedin \& Ostriker 1997, Abel \etal 1998a, Abel
\etal 1998b). Specifically in the context of CDM--type structure
formation models we have studied the collapse of high--$\sigma$
density fluctuations on small mass scales (Abel \etal~1998, herafter
AANZ98).

Here we present first results from three--dimensional adaptive mesh
cosmological hydrodynamics following the collapse and fragmentation of
the very first objects formed in hierarchical, CDM--like models of
structure formation.  

\section{Simulations}

The three dimensional adaptive mesh refinement calculations presented
here use for the hydrodynamic portion an algorithm very similar to the
one described by Berger and Collela (1989).  The code utilizes an
adaptive hierarchy of grid patches at various levels of resolution.
Each rectangular grid patch covers some region of space in its parent
grid needing higher resolution, and may itself become the parent grid
to an even higher resolution child grid. The general implimentation of
AMR places no restriction on the number of grids at a given level of
refinement, or the number of levels of refinement. Additionally the
dark matter is followed with methods similar to the ones presented by
Couchamn (1991). Furthermore, the algorithm of Anninos \etal (1997) to
solve the accurate time--dependent chemistry and cooling model for
primordial gas of Abel \etal (1997). Detailed description of the code
are given in Bryan \& Norman (1997,1998), and Norman \& Bryan
(1997,1998). 

The simulations are initialized at redshift 100 with density
perturbations of a sCDM model with $\Omega_B = 0.06$, $h=0.5$, and
$\sigma_8=0.7$. The abundances of the 9 chemical species (H, \Hp, \Hm,
He, \Hep, \Hepp, \HH, \HHp, e$^-$) and the temperature are initialized
as discussed in Anninos and Norman (1996). After a collapsing
high--$\sigma$ peaks has been identified in a low resolution
simulation the simulations is reinitialized with multiple refinement
levels on the Langrangian volume of the collapsing structure. The mass
resolution in the inital conditions within this Langrangian volume are
$0.53\Ms$ in the gas and $8.96\Ms$ for the dark matter component. The
refinement criterium ensures the local Jeans length to be resolved by
at least 4 grid zones as well as that no cell contains more than 4
times its initial mass of $0.53\Ms$. We limit the refinement to 12
levels within a $64^3$ top grid which translates to a maximum
dynamical range of $64\times 2^{12}=262,144$. As we will show below
the simulation is not resolution but physics limited.

\section{Results} 

\begin{figure}
{\centerline{\psfig{figure=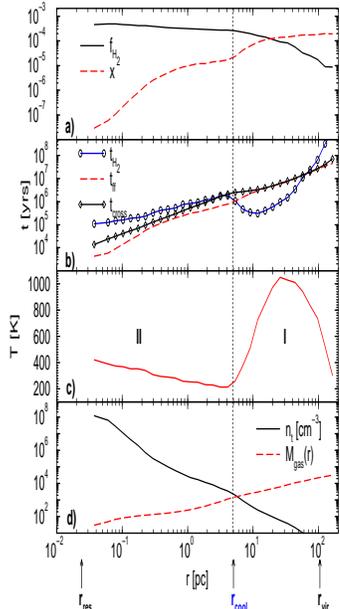,height=8cm,width=4.4cm}}}
\vspace{-0.4cm}
\caption[]{Spherically averaged mass weighted profiles around the
  baryon density peak at z=19.1. Panel a) shows the molcular hydrogen
  number fraction $f_{H_2}$ and the free electron number fraction $x$.
  Panel b) plots the \HH cooling time $t_{H_2}$, the time it takes a
  sound wave to travel to the center, $t_{cross}$, and the free--fall
  time $t_{ff}=[3\pi/(32G\rho)]^{1/2}$. Panel c) and d) show the
  temperature, baryonic number density and enclosed gas mass as a
  function of radius. The vertical dotted line indicates the cooling
  radius ($\sim 5pc$). The virial radius of the $5.6\tento{6}\Ms$ halo
  is $106\pc$. The cell size on the finest grid corresponds to
  $0.024\pc$. Note that the simulation box size corresponds to $6.4$
  proper $\kpc$.}
 \label{profile}
\end{figure}

In Figure~\ref{profile} we show mass weighted, spherically averaged
quantities around the densest cell found in the simulation at redshift
19.1 of various interesting quantities. In panel c) we indicate two
distinct regions. Region I) ranges from the virial radius to
$r_{cool}\sim 5\pc$, the radius at which the infalling material has
cooled down to $T\sim 200\K$. For most of this region the \HH cooling
time $t_{H2}$ is shorter than the free--fall time,
$t_{ff}=[3\pi/(32G\rho)]^{1/2}$, as is illustrated in panel b) of
Figure~\ref{profile}. The \HH number fraction rises from $7\tento{-6}$
to $2\tento{-4}$ as the free electron fraction drops from
$2\tento{-4}$ to $2\tento{-5}$ (panel a).

At $r_{cool}$ the free--fall time becomes smaller than the cooling
time. Also at $r_{cool}$ the time it takes a sound wave to travel to
the center, $t_{cross}=r/c_s=7.6\tento{6} r_{pc}/\sqrt{T_K}$\,yrs ,
becomes shorter than the cooling time. The cooling time approaches a
constant value at small radii (high densities) due to the transition
from non-LTE to LTE populations of the \HH rotational/vibrational
states.

\begin{figure*}
{\centerline{\psfig{figure=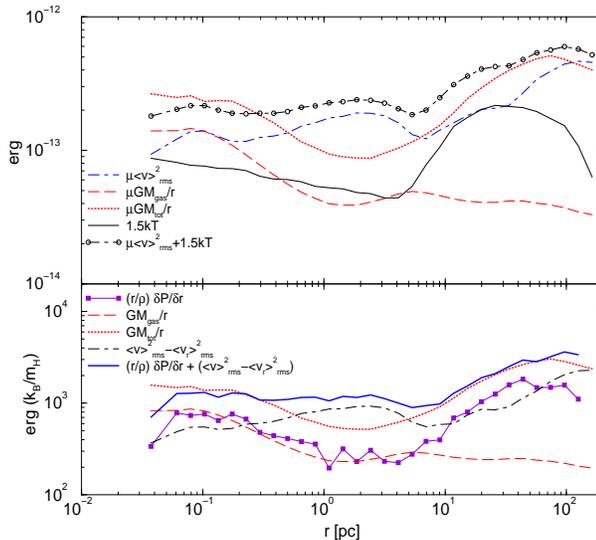,height=7.2cm,width=8.cm}}}
\vspace{-0.6cm}                 
\caption[]{Spherically averaged mass weighted profiles around the
  baryon density peak at z=19.1. The upper panel compares local
  gravitational, kinetic, and thermal energies. The lower panel
  compares centrifugal $([\langle v \rangle_{rms}^2- \langle v_r
  \rangle_{rms}^2]/r)$, pressure $(\rho^{-1} \partial P/\partial r)$,
  and gravity $(GM(r)/r^2)$ forces. They are muliplied with $r
  m_H/k_B$ for illustration purposes.  Within 0.3 pc gravitation
  dominates leading to a steeper density profile than in the region
  from 0.3 to 10 pc where the sum of centrifugal and pressure forces
  dominate.   }
 \label{fe_profile}
\end{figure*}

From the time scales one would conclude that below $r_{cool}$ the gas
should evolve quasi--hydrostatically on the cooling time scale. This
is, however, not strictly true. It turns out that centrigufal forces
play an important role in the collapse. To illustrate this
Figure~\ref{fe_profile} shows the mass weighted radial profiles of
various energies and force terms. From the lower panel of
Figure~\ref{fe_profile} it is evident that pressure + centrifugal
forces dominate gravity in the range of $0.3<r/1\pc< 10$ yielding a
net outward force (deceleration). Indeed the gas has zero radial
velocity at $r=0.3\pc$, separating the contracting core of $\sim
100\Ms$ at smaller radii.  There are multiple possible origins for a
``centrifugal barrier'' as the one observed at $r=0.3\pc$. The
simplest explanation being angular momentum conservation. However,
this would imply $v_\perp \times r =const.$ and the centrifugal force,
$F_c$, would be proportional to $r^{-3}$.  However, $F_c\simpr r^{-1}$
holds for more than three orders of magnitude, as can be seen from
Figure~\ref{fe_profile}. However, the cooling instability and the
continous merging of small scale structure offer simple alternative
explanations.

\section{Discussion}

Multiple interesting features of the collapsing and fragmenting
``primordial molecular cloud'' are identified. Most notably is the
fact of the existence of a contracting quasi--hydrostatic core of
$\sim 100\Ms$. Within this core the number densities increase from
$10^{5}$ to $10^8\cm^{-3}$. For densities $\gsim 10^6\cm^-3$, however,
three--body formation of \HH becomes the dominant formation mechanism
transforming all hydrogen into its molecular form (Palla \etal1983).
Our chemical reaction network does not include this reaction and the
solution cannot be correct at $r\lsim 0.1\pc$. The most interesting
effect of the tree--body reaction is that it will increase the cooling
rate by a factor $\sim 10^3$ leading to a further dramatic density
enhancement within the core. This will decrease the dynamical
timescales to $\ll 100\yrs$ effectively ``decoupling'' the evolution
of the fragment from the evolution of its ``host primordial molecular
cloud''.  Silk (1983) has argued that (due the enhanced cooling from
the 3--body produced \HH) fragmentation of this core might continue
until individual fragments are opacity limited (i.e. they become
opaque to their cooling radiation). Verifying this possibility will
have to await yet higher resolution simulations.

Clearly, it is the evolution of the $\sim 100\Ms$ core that could
participate in population III star formation. If it would form an open
star cluster with 100\% efficiency about $6\tento{63}$ UV photons
would be liberated during the average life time of massive star ($\sim
5\tento{7}\yrs$). This is about hundred times more than the $\sim
4\tento{61}$ hydrogen atoms within the virial radius. However, the
average recombination time $(nk_{rec})^{-1}\sim 5\tento{5}\yrs$ within
the virial radius is a factor 100 less than the average lifetime of a
massive star. Hence, one expects only very small or zero UV escape
fractions for these objects.

The \HH column density from the core to the virial radius is $\sim
10^{20}\cm^{-2}$. This is much larger than the typical
$10^{14}\cm^{-2}$ required for self--shielding in stationary
photodissociation regions (cf. Bertoldi and Draine 1996).
Hence also many sub--Lyman limit photons will be absorbed locally and
cannot reach the IGM. The host primordial molecular cloud is only
transparent for photons below $\sim 11\eV$ where no Lyman Werner Band
absorption can occur.

These optical depth arguments will almost certainly become incorrect
once supernova explosion alter the hydrodynamics and the chemistry and
cooling via collisional ionization, dissociation and shock heating.
More detailed understanding of the role of local feedback will have to
await new more detailed simulations.

\section{Conclusions}

We have reported first results from an ongoing project that studies
the physics of fragmentation and primordial star formation in the
cosmological context. The discussed results clearly illustrate the
advantages and power of structured adaptive mesh refinement
cosmological hydrodynamic methods to cover a wide range of mass,
length and timescales reliably. All findings of AANZ98 are confirmed
in this study. Among other things, these are that
\begin{itemize}
\item a significant number fraction of hydrogen molecules is only
  found in structures at the intersection of filaments
\item only a few percent of the gas in a virialized halo are cooled to
  $T\ll T_{vir}$.
\end{itemize}
The improvement of a factor $\sim 1000$ in resolution over AANZ98 has
given new ininsights in the details of the fragmentation process and
constraints on the possible nature of the first structures:
\begin{itemize}

\item Only $\lsim 1\%$ of the baryons within a virialized object can
  participate in population III star formation. 
\item The formation of super massive black holes or very massive
  objects in small halos seem very unlikely.
\item If the gas were able to fragment further through 3--body and/or
  opacity limited fragmentation only a small fraction of the gas
  will be converted into small mass objects.
\item The escape fraction of UV photons above the Lyman limit or in
  the Lyman Werner band should be small due to the high cloumn
  densities of HI ($N_{HI}\sim 10^{23}\cm^{-2}$) and \HH ($N_{H_2}\sim
  10^{20}\cm^{-2}$) in the surrounding of star forming regions. 
\end{itemize}

{\bf Cautionary Remark:} These latter conclusions are drawn from one
simulation and are to be understood as preliminary results from an
ongoing investigation.

\section*{Acknowledgments}
Tom Abel acknowledges support from NASA grant NAG5-3923 and useful
discussions with Karsten Jedamzik, Martin Rees, and Simon White.


\begin{thebibliography}{99}

\bibitem{Abel95} Abel, T. 1995, Thesis, University of Regensburg, Germany
\bibitem{Abel97a} Abel, T., Anninos, P., Zhang, Y.,  Norman, M. 1997a, 
        NewA, 2, 181
\bibitem{Abel98a} Abel, T., Anninos, P., Norman, M., Zhang, Y. 1998a, 
        ApJ, in press
\bibitem{Abel98b} Abel, T., Stebbins, A., Anninos, P., Norman,
    M.L. 1998b, ApJ, in press
\bibitem{Anninos96} Anninos, P., Norman, M.L. 1996, ApJ, 460, 556
\bibitem{Anninos97} Anninos, P., Zhang, Y., Abel, T.,  Norman,
  M.L. 1997, NewA, 2, 209
\bibitem{Berger89} Berger, M.J., Collela, P. 1989, J. Comp. Phys., 82, 64
\bibitem{Bertoldi96} Bertoldi, F., Draine, B. T. 1996, ApJ, 458, 222
\bibitem{Bodenheimer86}  Bodenheimer, P.H. 1986, {\sl Final Technical
    Report, California Univ., Santa Cruz.} 
\bibitem{Bryan97} Bryan, G.L., Norman, M.L. 1997, in {\it
        Computational Astrophysics}, eds. D.A. Clarke and M. Fall, ASP
        Conference \#123
\bibitem{Bryan98} Bryan, G.L., Norman, M.L. 1998, {\it in preperation}
\bibitem{Carr84} Carr,  B.J., Bond, J.R., Arnett, W.D. 1984, ApJ, 277, 445
\bibitem{Couchman91} Couchman, H. 1991, ApJL, 368, L23
\bibitem{Gnedin97} Gnedin, N.Y., Ostriker, J.P. 1997, ApJ, 486, 581
\bibitem{Haiman96} Haiman, Z., Thoul, A.A., Loeb, A. 1996, ApJ, 464, 523
\bibitem{Hirasawa69} Hirasawa, T. 1969, Progr. Theoret. Phys. , 42, 523
\bibitem{Hutchins76} Hutchins, J.B. 1976, ApJ, 205, 103
\bibitem{Kashlinsky83} Kashlinsky, A., Rees, M.J. 1983, MNRAS, 205, 955
\bibitem{MacLow86} Mac Low, M.-M. \& Shull, J.M. 1986, ApJ, 302, 585
\bibitem{Matsuda69} Matsuda, T., Sato, H., Takeda, H.1969,
        Progr. Theoret. Phys., 41, 840 
\bibitem{Norman97} Norman, M.L., Bryan, G.L. 1997, in {\it Workshop on
        Structured Adaptive Mesh Refinement Grid Methods}, ed. N. Chrisochoides 
\bibitem{Norman98} Norman, M.L., Bryan, G.L. 1998, in {\it Numerical
        Astrophysics 1998}, eds. S. Miyama \& K. Tomisaka
\bibitem{Palla83} Palla, F., Salpeter, E.E., Stahler, S.W. 1983, ApJ, 271, 632
\bibitem{Peebles68} Peebles, P.J.E.,  Dicke, R.H. 1968, ApJ, 154, 891
\bibitem{Puy96} Puy, D., Signore, M. 1996, A\&A, 305, 371
\bibitem{Silk1983} Silk, J. 1983, MNRAS, 205, 705
\bibitem{Takeda69} Takeda, H., Sato, H., Matsuda, T. 1969,
  Progr. Theoret. Phys., 41, 840
\bibitem{Tegmark97} Tegmark, M., Silk, J., Rees, M.J., Blanchard, A., Abel, T.,
        Palla, F. 1997, ApJ, 474, 1
\bibitem{Zhang97} Zhang, Y., Norman, M.L., Anninos, P., \& Abel,
        T. 1997, in  S.S. Holt and L.G. Mundy, eds., {\it Star
          formation, near and far}, AIP Press, New York, p329



\end{thebibliography}
\end{document}